\documentclass[conference]{IEEEtran}

\usepackage{graphicx}
\usepackage{pgfplots,amsmath,amssymb}
\usepackage{amsthm}
\usepackage{mathtools}
\usepackage{multirow}
\usetikzlibrary{shapes,arrows,fit,calc,positioning,automata}

\usetikzlibrary{pgfplots.fillbetween}

\pgfplotsset{
	tick label style={font=\small},
	label style={font=\small},
	legend style={font=\small}
}

\newtheorem{remark}{Remark}
 
\newlength{\figwidth}
\newlength{\figheight}
\newcommand{\rot}{\rotatebox{90}}
\newcommand{\argmin}[1]{\underset{#1}{\operatorname{arg}\,\operatorname{min}}\;}
\newcommand{\LD}{\ensuremath{l_\text{d}}}
\newcommand{\LE}{\ensuremath{l_\text{s}}}
\newcommand{\LamE}{\ensuremath{ \mathbf{\Lambda}_{\text{s}}}}
\newcommand{\EncS}{\ensuremath{ \varepsilon_\text{s}  }}
\newcommand{\Px}{\ensuremath{ \text{P}_{X} }}
\newcommand{\Pyx}{\ensuremath{ \text{P}_{Y|X} }}

\newcommand{\xone}{\ensuremath{  \mathcal{X}_{1}     }}
\newcommand{\xzer}{\ensuremath{  \mathcal{X}_{0}     }}

\definecolor{colb}{rgb}{0.0, 0.00, 0.7}%
\definecolor{colg}{rgb}{0.0, 0.50, 0.0}%
\definecolor{colr}{rgb}{0.7, 0.00, 0.0}%
\definecolor{colc}{rgb}{0.0, 0.50, 0.8}%
\definecolor{coly}{rgb}{0.8, 0.50, 0.0}%

\tikzstyle{uMLC}       = [dashed,line width=1.0pt,mark=o,mark options={solid},mark size=1.5pt]
\tikzstyle{SBSMLC}     = [solid,line width=1.0pt,mark=*,mark options={solid},mark size =1.5 ]
\tikzstyle{BICM}       = [dashed,line width=1.0pt,mark=square,mark options={solid},mark size =1.5 ]
\tikzstyle{sBICM}      = [solid,line width=1.0pt,mark=square*,mark options={solid},mark size =1.5 ]
\tikzstyle{RCB}        = [dash dot,line width=1.0pt]
\tikzstyle{Cap}        = [dotted,line width=1.0pt]

\begin{document}

\title{Sign-Bit Shaping Using Polar Codes}

\author{\IEEEauthorblockN{Onurcan \.{I}\c{s}can, Ronald B\"ohnke, Wen Xu}}

\maketitle

\begin{abstract}
A new polar coding scheme for higher order modulation is presented. The proposed scheme is based on multi-level coding (MLC) with natural labeling, where the bit-level corresponding to the sign-bit is generated in dependence on the previous bit-levels, such that the modulated symbols are distributed according to a target non-uniform distribution resulting in a shaping gain. 
This is realized by replacing the polar encoder in the sign-bit level by a successive cancellation (SC) decoder, such that the overall complexity is not increased compared to a conventional MLC scheme with polar codes.
Numerical simulations show significant performance improvements of the proposed approach compared to conventional transmission schemes with uniform symbol distribution. By using more complex decoders (e.g. SC list decoder), the proposed scheme outperforms Gallager's random coding bound.
\end{abstract}


\section{Introduction}

Polar coding \cite{Arikan09} can be regarded as a major breakthrough in coding theory, as it  is the first coding scheme with an explicit construction that provably achieves the channel capacity of binary input discrete memoryless channels. Beside the asymptotic optimality, its competitive performance in short and moderate block lengths (by using a list decoder \cite{Tal15}) made polar codes already a part of 5G New Radio (NR) \cite{chan_code5G}, the air interface technology used in the fifth generation mobile communication standard. 

After its invention, the ideas of polar coding have been employed for solving many information theoretic problems related to e.g. source coding \cite{hussami2009performance,korada2010polar}, broadcast channels \cite{goela2014polar}, asymmetric channels \cite{honda2013polar}, relay channels \cite{andersson2010nested}, information theoretic security \cite{andersson2010nested,6034749}. In addition, thanks to their nested structure, they are especially suitable  for applications, where problems need to be addressed jointly, such as joint channel coding and signal shaping \cite{shaped_polar}, or quantization and error correction \cite{8691487}. 

Similarly, polar codes are naturally suitable for transmission with higher order modulation, as they allow a joint description of coding and modulation \cite{seidl2013polar}. In this context, \cite{seidl2013polar} introduced polar coded modulation, where the symbol mapper is considered as an additional level of polarization. In that work, a multi-level coding (MLC) scheme with successive demapping (SD) (based on the  ideas from \cite{imai1977new,wachsmann1999multilevel}) is proposed. It turns out that MLC based approaches can achieve superior performance compared to bit-interleaved coded modulation (BICM) \cite{caire1998bit} based approaches with parallel demapping (PD). \cite{bocherer2017efficient} further discussed efficient code design for MLC based approaches.
Polar coding for BICM is analyzed from compound channels perspective in \cite{mahdavifar2015polar}, and was discussed in \cite{shin2012mapping}, where it is shown that mapping of the bits to symbols using a carefully designed interleaver (instead of a conventional random interleaver) and employing variable size kernels can improve the performance. A similar scheme is also discussed in \cite{chen2013efficient}, where $2\times2$ kernel (instead of variable size kernels) is used, and auxiliary virtual channels with zero-capacities are introduced to adapt different modulation orders. 

The mentioned works on polar codes considered coded modulation schemes with uniformly distributed symbols, which lead to a shaping loss. This loss can be up to 1.53dB on additive white Gaussian noise (AWGN) channels that can only be recovered if signal shaping is applied \cite{kschischang1993optimal}. Recently, signal shaping for higher order modulation got a renewed interest from the research community, and multiple approaches were presented on signal shaping for polar codes. A probabilistic shaping approach for polar codes is presented in \cite{shaped_polar}, where non-uniformly distributed Gray-labeled ASK symbols are generated (by extending the ideas of \cite{honda2013polar} and \cite{mondelli2018achieve}), and parallel demapping and a single stage decoding (similar to BICM) is employed. This approach is further extended to 5G NR polar codes in \cite{icscan2019probabilistic}. Another work is presented in \cite{lnt_hwdu} by combining the probabilistic amplitude shaping approach from \cite{bocherer2015bandwidth} with polar codes, where a precoder is used for systematic encoding and a distribution matcher is used to shape the amplitudes of the ASK symbols. The receiver uses a multi-stage decoder (as in MLC) and a distribution dematcher to recover the message.
Another MLC based approach is shown in \cite{icscan2018probabilistically}, where no distribution dematcher is required and where only a single bit-level is shaped to approximate the optimal symbol distribution. Another related work is presented in \cite{8492454}, where polar lattices are shown to be capable of achieving the AWGN capacity.


\subsection{Contributions}
In this work, we first analyze PS approaches in dependence on the demapping method (i.e. successive or parallel), and discuss their influence on the choice of the shaping parameters. We evaluate their  achievable rates, and show that the shaping rate need to be adjusted for the operating signal-to-noise ratio (SNR) for parallel demapping, whereas a fixed shaping rate is enough to compensate most of the shaping loss for a large SNR range when successive demapping is employed. We further find that shaping only a single bit-level is sufficient to obtain good results, and propose using sign-bit shaping with successive demapping.

Based on our findings, we show how this can be implemented in practice, and propose a \textit{sign-bit shaped multi-level coding} (SBS-MLC) scheme based on polar codes, which has the following advantages compared to existing polar coding approaches with probabilistic shaping:

\begin{itemize}
\item Compared to \cite{lnt_hwdu}, the proposed method does not require systematic encoding (which is realized with a precoder in \cite{lnt_hwdu}) and a separate shaping encoder (i.e. distribution matcher) at the transmitter. Moreover, a shaping decoder (i.e. distribution dematcher) is not required either, because the channel decoder at the receiver already outputs the message.

\item Compared to \cite{shaped_polar}, the proposed scheme uses an MLC based approach, and therefore any choice of the modulation order $m$ is possible. In \cite{shaped_polar}, however, it is assumed that $m$ is an integer power of $2$, and therefore 8-ASK ($m=3$) cannot be supported directly.   \cite{icscan2019probabilistic} solves this problem by modifying 5G NR polar codes and shaping only a single bit-level, where only a coarse approximation of the target distribution can be achieved. The presented approach in this work can more exactly approximate the target distribution.

\item Similarly, a single bit-level is shaped in \cite{icscan2018probabilistically} using an MLC approach. Unlike this work, \cite{icscan2018probabilistically} encodes each bit-level independently, resulting in a coarse approximation of the target distribution. The proposed scheme in this work also performs signal shaping on a single bit-level, however since the sign-bit is encoded depending on the previous bit-levels, a much more exact approximation of the target distribution can be obtained even for simpler implementations of the shaping encoder (e.g. by using an successive cancellation (SC) decoder instead of a  list decoder for shaping).
\item In \cite{8492454}, a Gaussian shaping approach is used to build lattice codes with polar coding ideas. Unlike our approach, \cite{8492454} requires all bit-levels to be encoded in dependence on the previous levels, and it requires a common randomness between the transmitter and the receiver.
\item SBS with convolutional codes was originally introduced in \cite{forney1992trellis} as a special case of trellis shaping based on lattice partitions, and it is further discussed in \cite{smith2012pragmatic}. In this work, we extend these ideas to polar codes, such that all bit-levels use polar coding. Moreover, our proposal allows a more flexible way to allocate shaping redundancy and coding redundancy compared to convolutional codes.
\end{itemize}

\subsection{Outline}
In Sec. \ref{sec:sysmod}, we introduce the system model and describe the transmission strategies. In Sec. \ref{sec:theo}, we discuss the theoretically achievable rates and elaborate on signal shaping for different strategies. 
In Sec. \ref{sec:polar} we describe polar coding based MLC and show how sign-bit shaping can be implemented efficiently using polar codes. In Sec. \ref{sec:PerfEval}, performance evaluations on AWGN channels are given.
Sec. \ref{sec:Conc} concludes the paper.

\subsection{Notation}

In this work, we use lowercase bold letters (e.g. $\mathbf{x}$) for vectors, uppercase letters (e.g. $X$) for random variables representing elements of the associated vectors and lowercase letters for their realizations. $\Px$, $\text{H}(X)$ and $\text{E}(X)$ denote the probability distribution, entropy and expected value of $X$, respectiveley. $\text{I}(X;Y)$ is the mutual information between $X$ and $Y$.  Calligraphic letters (e.g. $\mathcal{F}$) represent sets.

\section{System Model}
\label{sec:sysmod}
Consider the AWGN channel model 
\begin{align}
\mathbf{y} = \mathbf{x} + \mathbf{z},
\end{align}
where $\mathbf{z}$ contains the Gaussian noise with variance $\sigma^2$, $\mathbf{x}$ contains $2^m$-ASK channel input symbols taken from the alphabet 
\begin{align}
\mathcal{X}=\{\pm 1, \pm 3,...,\pm(2^m-1)  \}.
\end{align}
The SNR becomes $\text{E}(X^2)/{\sigma^2}$. Note that the SNR depends on $\Px$, and for uniform distribution it simplifies to $(2^{2m}-1)/3\sigma^2$.
Below we discuss different transmission strategies.

\subsection{Multi-Level Coding}
\begin{figure}
	\centering
	\includegraphics{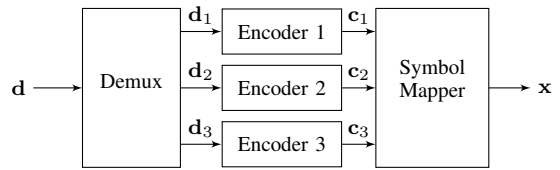}	
	\caption{MLC transmitter with $m=3$ bit-levels.}
	\label{fig:BlcDiagram_MLC}
\end{figure}

In multi-level coding approach, the message $\mathbf{d}$ of length $k$ is divided into $m$ parts $\mathbf{d}_i$ of lengths $k_i$  with $k=\sum_{i}{k_i}$.
Each of them is then encoded separately to codewords $\mathbf{c}_i$ of lengths $n_c$. A symbol mapper uses one bit from each of the codewords to generate a $2^m$-ASK symbol, i.e. each codeword is mapped to a specific ASK bit-level. The transmission rate is $R_\text{MLC}=k/n_c$.
Fig. \ref{fig:BlcDiagram_MLC} depicts the block diagram of the transmitter with $m=3$. 

\begin{figure}
	\centering
	\includegraphics{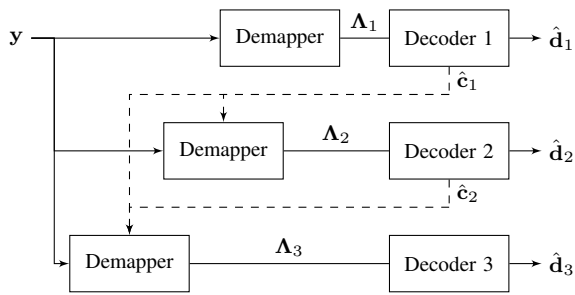}	
	\caption{Multistage demapper/decoder for MLC transmission.}
	\label{fig:BlcDiagramRec}
\end{figure}

The receiver performs successive demapping and multi-stage decoding to obtain the estimate $\hat{\mathbf{d}}_i$ as depicted in Fig. \ref{fig:BlcDiagramRec}. 
At stage $i\in\{1,\cdots, m\}$, bit log-likelihood ratios (LLRs) $\mathbf{\Lambda}_i$ corresponding to the $i$th ASK bit-level are obtained using a symbol demapper and are decoded to obtain $\hat{\mathbf{c}}_i$. The demapper uses the estimate $\hat{\mathbf{c}}_j$, $j\in\{1,\cdots, i-1\}$ from previous stages as a-priori information to remove the dependencies between bit-levels successively. More specifically, the $i$th bit LLR is calculated from a received symbol $y$ as 
\begin{align}
\label{eq:BitLLR_s}
\Lambda_i &= \log \frac{\sum\limits_{x_j\in \xzer}  \Pyx(y|x_j) \Px(x_j) }{\sum\limits_{x_j\in\xone}  \Pyx(y|x_j) \Px(x_j) }. 
\end{align}
Here, $\xzer$ and $\xone$ are subsets of $\mathcal{X}$ containing symbols, whose bits at positions $j\in\{1,\cdots, i-1 \}$ are $\hat{c}_j$, and the bit at position $i$ is a $0$ or $1$, respectively.

\subsection{Bit-Interleaved Coded Modulation}
\begin{figure}
	\centering
	\includegraphics{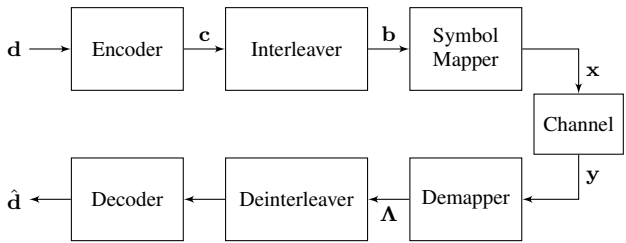}	
	\caption{BICM transmission chain.}
	\label{fig:BlcDiagram_bicm}
\end{figure}

In bit-interleaved coded modulation, the message $\mathbf{d}$ of length $k$ is encoded into a single codeword $\mathbf{c}$ of length $m n_c$, which is interleaved and mapped to $2^m$-ASK symbols $\mathbf{x}$, as depicted in Fig. \ref{fig:BlcDiagram_bicm}. Usually, Gray labeling is preferred for symbol mapping. The transmission rate is $R_\text{BICM}=k/n_c$.
At the receiver, parallel demapping of bit-levels  is used, followed by the deinterleaver and the decoder to obtain $\hat{\mathbf{d}}$.
Note that the bit LLRs $\Lambda_i$ with PD can be calculated similar to (\ref{eq:BitLLR_s}), except $\xzer$ and $\xone$ are defined only in dependence on $i$, and not on other bits. This allows to reduce the receiver latency compared to SD (since all bit-levels can be demapped in parallel), but is suboptimal as it neglects the dependencies between bit-levels.


\section{Theoretical Evaluations}
\label{sec:theo}
In this section, we briefly discuss the achievable rates of different transmit schemes as a motivation for our proposed method in Sec. \ref{sec:polar}.

\subsection{Comparison of Achievable Rates}

The channel capacity corresponds to the maximum data rate that allows for reliable transmission with arbitrarily small error probability. For an AWGN channel with average transmit power constraint, it is given by
\begin{align}
\label{eq:Capacity}
C_{\text{AWGN}} = \max_{\Px :\, \text{E}(X^2) \leq P} \text{I}(X; Y) 
= \frac{1}{2} \log_2(1 + \text{SNR}) \;.
\end{align} 
The optimal continuous Gaussian input distribution can be well approximated in practice by ASK symbols drawn from a Maxwell-Boltzmann (MB) distribution 
\begin{align}
\label{eq:MBdist}
\Px(x) = \dfrac{\exp(-\nu x^2)}{\sum\limits_{x_i\in\mathcal{X}} \exp(-\nu x_i^2) }.
\end{align}
According to the chain rule of mutual information \cite{gallager1968information}, $\text{I}(X;Y) = \text{I}(C_1,\ldots,C_m ; Y)$ can be achieved using MLC with successive demapping
\begin{align}
\label{eq:Rsd}
R_{\text{SD}}  &= \sum_{i=1}^m \text{I}(C_i; Y| C_1,\cdots  C_{i-1})
\;\nonumber\\
 &= \text{H}(X) - \sum_{i=1}^m \text{H}(C_i| Y, C_1,\cdots  C_{i-1}).
\end{align}
This asymptotic rate does not depend on the symbol mapping, but natural binary labeling is superior for finite block lengths \cite{wachsmann1999multilevel}.

With parallel demapping of the bits without conditioning on previous decisions, non-negative rates
\begin{align}
\label{eq:Rpd}
R_{\text{PD}} &= \text{H}(X) - \sum_{i=1}^m \text{H}(C_i| Y)
\end{align}
can be achieved \cite{bocherer2014achievable}. The loss due to neglecting the dependencies of the bit-levels in the demapper is in general minimized by Gray labeling, where neighboring symbols differ in only one bit. For BICM, the bits $C_1,\ldots,C_m$ are independently distributed such that $\text{H}(X) = \sum_i{\text{H}(C_i)}$, which leads to \cite{caire1998bit}
\begin{align}
\label{eq:Rbicm}
R_{\text{BICM}} = \sum_{i=1}^m \text{I}\left(C_i; Y\right).
\end{align}

\begin{remark}
The second terms in (\ref{eq:Rsd}) and (\ref{eq:Rpd}) can be seen as the coding redundancy, i.e. redundancy introduced by the channel code. Similarly, $m-\text{H}(X)$ corresponds to the shaping redundancy, or the shaping rate $R_s$.
\end{remark}

\begin{figure}
	\centering
	\includegraphics{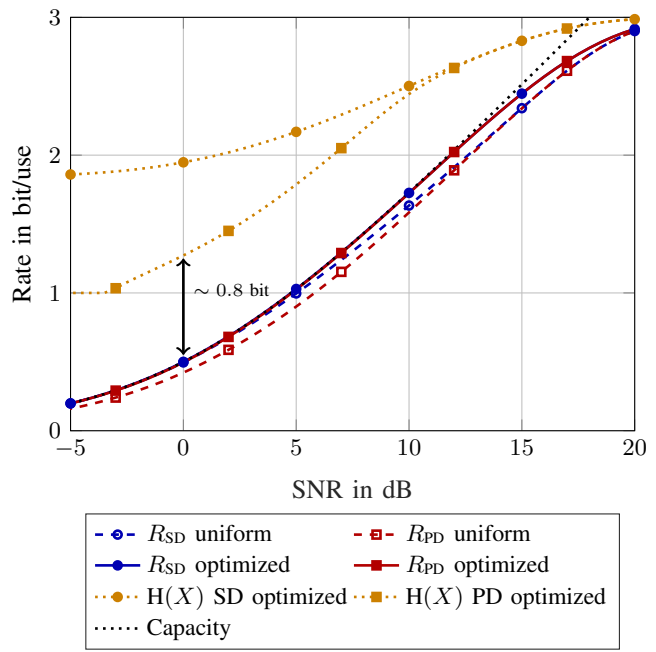}	
	\caption{Achievable rates for 8-ASK with optimized and uniform symbol distributions.}
	\label{fig:AchRates}
\end{figure}

Fig. \ref{fig:AchRates} compares the achievable rates for 8-ASK. For uniformly distributed transmit symbols, there is a shaping loss at high SNR and an additional demapping loss for PD with Gray labeling at low SNR. Using an MB distribution $\nu$ optimized for the operating SNR, however, the channel capacity can be closely approached with both successive and parallel demapping (the curves overlap in Fig. \ref{fig:AchRates}). 

We observe that the gap between $R_\text{PD}$ and the corresponding optimal entropy $\text{H}(X)$ is almost constant over a wide SNR range, which suggests to have a constant coding redundancy of approximately $0.8$ bits per symbol, and to adjust the data rate through shaping as in \cite{bocherer2015bandwidth}. For SD, on the other hand, a constant fraction of the total redundancy should be dedicated to coding and shaping, respectively \cite[Sec. VIII]{wachsmann1999multilevel}. This means that depending on the choice of the demapping method, one should allocate shaping and coding redundancy differently.

\subsection{Achievable Rates with Fixed Input Distribution}
\label{sec:FixDist}
\begin{figure}
	\centering
	\includegraphics{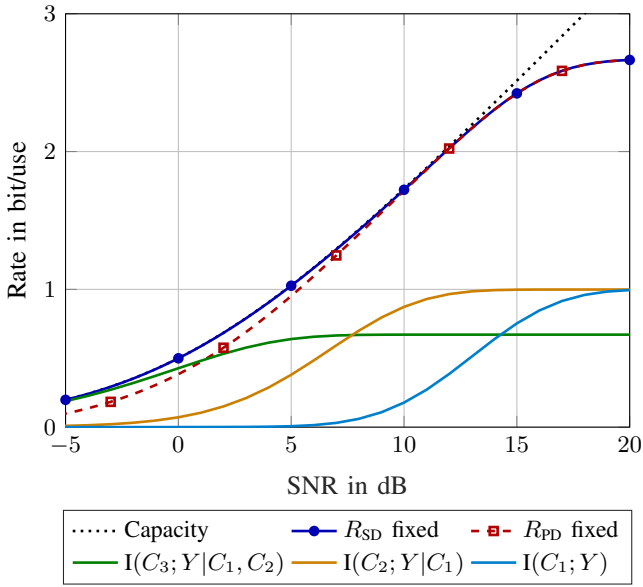}	
	\caption{Achievable rates for 8-ASK with fixed MB distribution having $\text{H}(X) = 2.66\;\text{bit}$.}
	\label{fig:MLCRates}
\end{figure}

Fig. \ref{fig:MLCRates} shows the achievable rates for 8-ASK using a fixed MB distribution with entropy $\text{H}(X) = 2.66\;\text{bit}$, corresponding to $R_s \approx 1/3$. We observe that $R_\text{SD}$ is very robust with respect to the input distribution, as the capacity is closely approached over a wide SNR range. In contrast, the gap to capacity increases for $R_\text{PD}$ at lower rates due to the demapping loss.

Fig. \ref{fig:MLCRates} also includes the rates $\text{I}(C_i; Y| C_1,\cdots C_{i-1})$ of the bit-levels for natural labeling. These rates converge to one at high SNR for the first two bit-levels, which means that $\text{H}(C_i| C_1,\cdots C_{i-1}) \approx 1$, and hence the corresponding bits must be approximately uniformly distributed. Thus, an almost optimal transmit symbol distribution can be obtained by shaping only the sign bit $C_3$. We observed similar results for other modulation orders.

Note that in order to obtain an MB distribution with smaller entropy $\text{H}(X)$ according to Fig. \ref{fig:AchRates}, it is in general required to shape multiple bit-levels, which increases the implementation complexity at the transmitter. On the other hand, a smaller shaping rate would lead to a more uniform symbol distribution and hence a larger shaping loss.
Sign-bit shaping with $R_s \approx 1/3$ represents a good tradeoff between performance and complexity.

\begin{remark}
If shaping for PD with Gray labeling is considered, one can use a fixed coding redundancy and adjust the transmission rate through the symbol distribution. This is advantageous, if the rate adaptation through channel code is cumbersome, and flexible shaping encoders are available.
On the other hand, for SD one can use a fixed shaping redundancy (a fixed shaping code), and adjust the transmission rate through the channel code itself. This is advantageous, if the used channel code is flexible in terms of rate adaptation, and the shaping encoder is not necessarily flexible in terms of supporting different rates.
\end{remark}

\subsection{Design Rules}
Based on the above discussion, we propose to use MLC with a fixed shaping rate for ASK with natural labeling. According to Fig. \ref{fig:AchRates}, $R_\text{SD}$ starts deviating from the capacity for rates approximately above $(m-1)$ bit/use due to the limitation of the finite constellation size. Furthermore, $\text{I}(C_1; Y)$ goes to zero in Fig. \ref{fig:MLCRates} for rates below $(m-2)$ bit/use, so this bit-level does not carry any information. Consequently, $2^m$-ASK should be applied preferably for rates within $(m-2,m-1]$ bit/use. In this range, we found that an input distribution with entropy $\text{H}(X) \approx (m - 1/3)\;\text{bit}$ is close to optimal. Hence, approximately $1/3$ of the sign bits will be used for shaping the input distribution. Note that only the first few bit-levels need to be protected against errors for successive demapping of natural labeling due to the increasing Euclidean distances in each step, so uncoded transmission may be used in the last bit-levels for large values of $m$.

\section{Sign-Bit Shaped Multi-Level Polar Codes}
\label{sec:polar}

In this section, we discuss how the design rules obtained in the previous section can be applied to polar codes. 

\subsection{Polar Codes}
Polar coding \cite{Arikan09} relies on the \textit{channel polarization} phenomenon, where the physical channel is converted into polar sub-channels, which tend to have very high or very low reliabilities asymptotically. A polar encoder assigns message bits to reliable channels, and (known) frozen bits to unreliable channels. A polar decoder (such as an SC decoder) processes a noisy observation of the polar codeword together with the frozen bits to estimate the message bits.

Let $\mathbf{G}$ denote the polar transform matrix of size $n_c \times n_c$, which is defined as the $(\log_2 n_c)$-th Kronecker power of the $2\times2$ kernel $\begin{psmallmatrix}1 & 0\\ 1 & 1\end{psmallmatrix}$.
A polar codeword $\mathbf{c}$ is obtained from the input sequence $\mathbf{u}$ by
\begin{align}
\label{eq:polareq}
\mathbf{c} = \mathbf{u}\mathbf{G}.
\end{align}
Here, $\mathbf{u}$ contains the message bits $\mathbf{d}$ at indices $\mathcal{I}$, and frozen bits \textit{}at indices $\mathcal{F}$ with $\mathcal{I}\cup\mathcal{F}=\{1,\cdots,n_c\}$, where $\mathcal{I}$ and $\mathcal{F}$ denote the sets containing the indices of sub-channels with high and low reliabilities, respectively. The performance of a polar code depends on the choice of these sets. For a given channel, one can obtain reliabilities of polar sub-channels using density evolution \cite{mori2009performance, Tal13} or its approximations \cite{trifonov2012efficient}, and use the most reliable $k$ sub-channels for $\mathcal{I}$, and the other $n_c-k$ sub-channels for $\mathcal{F}$ to obtain a code with rate $k/n_c$.

Fortunately, the order of the sub-channel reliabilities is similar for many symmetric channels. In 5G NR, a polar sequence $\mathbf{Q}$ is specified that is used to obtain $\mathcal{I}$ and $\mathcal{F}$ in a simple way \cite{chan_code5G}. Accordingly, the first $n_c-k$ indices and the last $k$ indices in $\mathbf{Q}$ that are less than $n_c$ are used for $\mathcal{F}$ and $\mathcal{I}$, respectively. This allows a very simple and flexible code design with relatively good performance.
In the rest of this work, we will also use the polar sequence $\mathbf{Q}$ from 5G NR to design our codes.

It is known that the SC decoder performs well only for very long codeword lengths, and the SC list (SCL) decoder \cite{Tal15} improves the performance significantly for shorter lengths. An SCL decoder operates similarly to an SC decoder, but does not make a decision for each bit directly. Instead, it considers multiple decoding paths concurrently at each decoding step, and outputs a list of candidate codewords together with their path metrics (PM), which is related to the a-posteriori probability of each candidate. Picking the codeword with the best PM from the list already improves the decoding performance of polar codes, but a further improvement can be obtained with a cyclic redundancy check (CRC) aided list decoding \cite{Tal15}, where an outer CRC code is used to select the correct codeword from the decoder output. Although this effectively reduces the number of frozen bits (due to the additional CRC bits), the decoding performance improves compared to SCL decoders with only PM based codeword selection.

\subsection{Multi-Level Coding Using Polar Codes}
\label{sec:MLC}
Polar codes have certain advantages that are well suited for MLC.
Firstly,  MLC approaches require multiple channel codes (of different rates) that have the same codeword length. Polar codes naturally allow a very flexible rate adaptation for a given codeword length $n_c$.
Secondly, thanks to their structure, SC and SCL decoders can already output the codeword (beside the message bits) that is required for demapping the next bit-levels. In this way, no additional encoding of the message bits at the receiver is required.
Last but not least, SCL decoders can be initialized with a list based on the output of the previous decoding stages. This is advantageous, because it allows passing information between decoding stages that we discuss below.

Since MLC approaches use multiple short channel codes (instead of a single long channel code as in BICM), a larger finite length loss may be expected. On the other hand, one can consider the symbol mapping as an additional level of polarization \cite{seidl2013polar} that combines $m$ short polar codewords of length $n_c$ to a long codeword of length $m n_c$. Note that natural labeling causes the bit-level capacities to have a large variance \cite{wachsmann1999multilevel,seidl2013polar}, i.e. it is a good polarizing kernel. With this fact in mind, we propose using a \textit{list multi-stage decoder} with SD similar to \cite{dettmar1992new} and \cite{lnt_hwdu}. Accordingly, the receiver performs the following steps:

\begin{itemize}
	\item The received noisy sequence $\mathbf{y}$ is demapped to obtain bit LLRs $\mathbf{\Lambda}_1$ for the first ASK bit-level as in conventional MLC approaches (as in Fig. \ref{fig:BlcDiagramRec}), which is processed using a list decoder to obtain $\LD$ candidate codewords $\mathbf{c}_{1,j}$ and their path metrics $\text{PM}_j$ with $j\in\{1,\cdots, \LD \}$. 
	\item $\mathbf{y}$ is demapped separately using each candidate $\mathbf{c}_{1,j}$ as a-priori information, and  $\LD$ bit LLR sequences $\mathbf{\Lambda}_{2,j}$ for the second ASK bit-level are obtained.
	\item A list decoder is used to obtain $\mathbf{c}_{2,j}$, where the list is initialized with $\LD$ sequences $\mathbf{\Lambda}_{2,j}$. Similarly, the path metrics are initialized with $\text{PM}_{j}$. The output contains $\LD$ estimates $\mathbf{c}_{2,j}$. The decoder also keeps the relation $\mu_{2,j}$ between the input and output sequences, i.e. an index for each of the output sequences $\mathbf{c}_{2,j}$, that indicates to which sequence in the initial list its decoding path belongs to. 
	\item The same procedure is applied for the remaining bit-levels. After the final bit-level is processed, the candidate codewords from each stage can be extracted using $\mu_{i,j}$ and $\mathbf{c}_{i,j}$.
\end{itemize}
Note that a conventional SCL decoder is initialized with a single path, and during the decoding process the number of paths are increased. In the presented scheme, the only  modification to the decoder is that it is initialized with $\LD$ paths and their corresponding path metrics.
This procedure can also be seen as multi-kernel polar decoding, where the operation in the first polarization step is performed using a symbol demapper. 
Accordingly, the list multi-stage decoder allows decoding $m$ polar codewords of length $n_c$ treating them as a single codeword of length $mn_c$. Therefore, any error made at stage $i$ can be compensated by the later stages, provided that the correct codeword is in the output list of the $i$th decoder. 

This scheme also allows an outer CRC-code, which can be used to select the correct candidate from the list in the last stage (instead of using a separate CRC outer code for each ASK bit-level).

\subsection{Sign-Bit Shaping with Polar Codes}

In Sec. \ref{sec:FixDist} we have shown that shaping a single bit-level with MLC can be enough to compensate for the shaping loss. In this section we propose a polar coding based transmitter and receiver that we call \textit{sign-bit shaped MLC} (SBS-MLC).

\begin{figure}
	\centering
\includegraphics{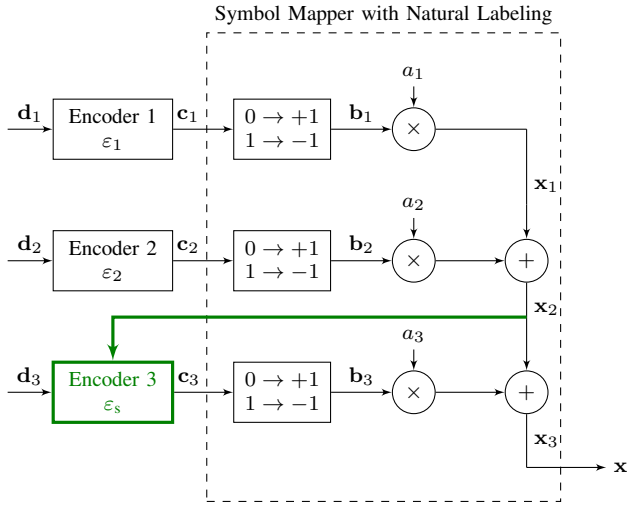}	
	\caption{Block diagram of the proposed transmitter.}
	\label{fig:BlcDiagram_SBS}
\end{figure}

\subsection*{Transmitter}

\begin{figure}
	\centering
	\includegraphics{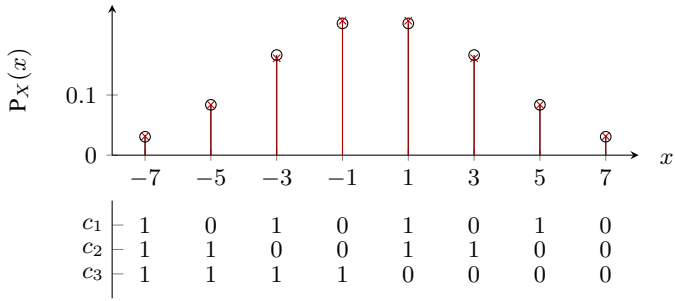}	
	\caption{Natural labeling for 8-ASK according to the symbol mapper in Fig. \ref{fig:BlcDiagram_SBS}, and the resulting $\text{P}_X$, if an SC decoder with min-sum approximation is used for $n_c=256$ and $s=84$ (circle markers). The reference $\text{P}_X$ ($\times$ markers) shows the MB distribution with the same average energy.}
	\label{fig:pmf}
\end{figure}

Fig. \ref{fig:BlcDiagram_SBS} depicts the proposed transmitter with natural labeling for $8$-ASK (with $m=3$ bit-levels), which can easily be extended to any choice of $m$. Here, the $8$-ASK symbols are represented by the superposition of three 2-ASK symbols that are weighted by $a_i=2^{i-1}$. Observe that the most significant bit ($i=3$) is the sign bit. Fig. \ref{fig:pmf} shows the resulting labeling.

In the proposed scheme, the first $m-1$ stages are equivalent to stages of a conventional MLC encoder, where codewords with uniformly distributed bits are generated using polar encoders $\varepsilon_1$ and $\varepsilon_2$. In the last stage, however, we use a modified encoder $\EncS$ to obtain the codeword $\mathbf{c}_3$ for the sign bit-level.
The task of this encoder is to generate $\mathbf{c}_3$ in dependence on $\mathbf{c}_1$ and $\mathbf{c}_2$, such that the resulting ASK symbols $\mathbf{x}$ at the output of the symbol mapper have a target probability distribution as in (\ref{eq:MBdist}), which is optimal for AWGN channels. 

Observe the relations from Fig. \ref{fig:BlcDiagram_SBS}
\begin{align}
\mathbf{x}_i &= \sum_{j=1}^{i} a_j \cdot (-2\cdot \mathbf{c}_j + 1)\\
 &= \mathbf{x}_{i-1} +  a_i \cdot (-2\cdot \mathbf{c}_i + 1).
\end{align}
Accordingly, by generating $\mathbf{c}_3$ (i.e. the codeword in the sign bit-level) in dependence on $\mathbf{x}_2$, we can force the ASK symbols in $\mathbf{x}$ to have a target probability distribution $\Px$. For this purpose, we allocate the most reliable $s$ polar sub-channels in the sign bit-level for \textit{shaping bits}, which do not carry any information, but cause $\mathbf{x}$ to have the desired probability distribution. 
As a result, obtaining the bits in the sign bit-level can be formulated as a channel decoding problem, i.e. the encoder $\EncS$ in the sign bit-level is actually a \textit{polar decoder} that uses $\mathbf{x}_2$ to search for a codeword representing $\mathbf{d}_3$ and at the same time causing $\mathbf{x}$ to be distributed according to the target $\Px$.   

In light of these facts, we formulate the operations of obtaining $\mathbf{c}_3$ as a polar decoding problem (similar to \cite{shaped_polar}), where we treat
\begin{itemize}
	\item the message bits $\mathbf{d}_3$ (of length $k_3$) as frozen bits at indices $\mathcal{I}$,
	\item an all-zero vector of length $n_c-k_3-s$ as frozen bits at indices $\mathcal{F}$, 
	\item polar sub-channel indices given in $\mathcal{S}$ as the indices of $s$ unknown bits to be recovered by the decoder,
	\item and $\LamE$ (defined below) as the decoder input in LLR form.
\end{itemize}
Here, assuming a fixed polar sequence, $\mathcal{S}$ contains the indices of the most reliable $s$ sub-channels, $\mathcal{I}$ contains the indices of the remaining most reliable $k$ sub-channels, and $\mathcal{F} = \{1,\cdots n_c \} \setminus (\mathcal{S}\cup \mathcal{I})$.  Note that this is equivalent to attaching the $s$ shaping bits to $\mathbf{d}_3$, and using a polar encoder with rate $(k_3+s)/n_c$ (instead of $k_3/n_c$). Since a polar decoder can already output the codeword, this additional encoding operation is not required.

Observe that $x = x_2+a_3$ if $c_3=0$, and $x = x_2-a_3$ otherwise. Accordingly, we can define 
\begin{align}
\Lambda_3 &= \log \frac{\text{P}_{C_3}(0)}{\text{P}_{C_3}(1)} =  \log\frac{\Px(x_2 + a_3)}{\Px(x_2 - a_3)} \nonumber\\
&= -\nu (x_2 + a_3)^2 + \nu (x_2 - a_3)^2 \nonumber \\
&= -4\cdot \nu \cdot a_3 \cdot x_2
\end{align}
as the LLR value for $c_3$ as a function of $x_2$, assuming an MB distribution with parameter $\nu$ is targeted.
As a result, 
\begin{align}
\LamE = -4\cdot \nu \cdot a_3 \cdot \mathbf{x}_2
\end{align}
should be used as the decoder input in LLR form for $\varepsilon_s$. For any choice of $m\geq2$, this corresponds to  
\begin{align}
\LamE = - \nu \cdot 2^{m+1} \cdot \mathbf{x}_{m-1}.
\end{align}

Note that a hardware friendly implementations of an LLR based polar decoder may use the min-sum approximation to simplify the check-node operations \cite{B-Stimming15}. This approximation makes the decoder output independent of the scaling of the LLR input. 
Therefore, we can simply formulate $\LamE$ as 
\begin{align}
\LamE = -\mathbf{x}_{m-1},
\end{align}
if a decoder with min-sum approximation is employed. This means that the decoder $\EncS$ does not even need to know the exact $\Px$ to generate $\mathbf{c}_{m}$.

\begin{remark}
The problem of obtaining $\mathbf{c}_m$ can also be formulated as an energy minimization problem of $\mathbf{x}$.
\begin{align}
\label{eq:ml}
\mathbf{c}_m &= \argmin{\mathbf{c}} ||\mathbf{x} ||^2 \nonumber\\
&= \argmin{\mathbf{c}} ||  a_m\cdot (-2\cdot\mathbf{c} + 1) + \mathbf{x}_{m-1} ||^2.
\end{align}
Observe that (\ref{eq:ml}) can also be interpreted as the maximum likelihood solution for transmission of a BPSK modulated binary sequence $\mathbf{c}_m$ over an AWGN channel with channel gain $a_m$ \cite{tse2005fundamentals}. In this case, $-\mathbf{x}_{m-1}$ can be seen as the received noisy sequence, which should be the input of the maximum likelihood decoder. Also note that the MB distribution minimizes the average energy for a fixed entropy \cite{kschischang1993optimal}. This is another reason, why the energy minimization problem given above would result in $\mathbf{x}$ with an MB distribution. 
\end{remark}

Recall that reserving roughly $1/3$ bit per symbol for shaping is a good choice to approach channel capacity for a wide SNR range. 
Fig. \ref{fig:pmf} shows the resulting distribution of 8-ASK symbols (obtained by simulations) if an SC decoder with min-sum approximation is used with $s=84$ shaping bits to generate $n_c=256$ ASK symbols ($R_s\approx 1/3$), and where $\mathcal{S}$ is constructed from the most  reliable $s$ indices in $\mathbf{Q}$ from the 5G NR specification \cite{chan_code5G}.

\begin{remark}
The obtained pmf has an entropy of $2.728$ bits, although theoretically $m-R_s=2.671$ should be possible for $R_s=84/256$. This means that the SC decoder and the finite block length cause a rate loss of $0.057$ bits per channel use, corresponding to approximately 15 additional shaping bits for $n_c=256$. 
\end{remark}

Note that an SC decoder and a polar encoder have the same order of complexity $O(n_c \log n_c)$ \cite{Arikan09}. Moreover, as $\LamE$ consists of only integers and as the min-sum approximation results in simplified decoder operations, the complexity increase due to replacing a polar encoder with an SC decoder at the transmitter is small. On the other hand, if the transmitter has more computational power, one can also use an SCL decoder as $\EncS$, which results in a better performance at the cost of increased complexity. In the example above with an SC decoder (given in Fig. \ref{fig:pmf}), the average transmit power corresponds to $\sum_i x_i^2 \cdot \text{P}_X (x_i) = 10.26\text{dB}$. If an SCL decoder (using min-sum approximation) with list size $\LE=32$ is used at the transmitter, the average transmit power would become 10.06dB, resulting in a 0.2dB gain in the SNR without an additional complexity at the receiver.

\subsection*{Receiver}
A conventional MLC receiver with two modifications can be used at the receiving side.

\subsubsection*{Demapping}
For an AWGN channel with noise variance $\sigma^2$ and uniformly distributed $X$, the bit LLRs given in  (\ref{eq:BitLLR_s}) can be written as 
\begin{align}
\label{eq:BitLLR_uniform}
\Lambda_i^{\text{uni}} = \log \frac{\sum\limits_{x_j\in\xzer}  \exp\left( \frac{ -(y-x_j)^2 }{ 2\sigma^2} \right)  }{\sum\limits_{x_j\in\xone}  \exp\left( \frac{ -(y-x_j)^2 }{ 2\sigma^2} \right)  }.
\end{align}
Similarly, for the MB distribution given in (\ref{eq:MBdist}), we can show that (\ref{eq:BitLLR_s}) simplifies to
\begin{align}
\label{eq:BitLLR_MB}
\Lambda_i^{\text{MB}} = \log \frac{\sum\limits_{x_j\in\xzer}  \exp\left( \frac{ -(\alpha y-x_j)^2 }{ 2 \alpha  \sigma^2} \right)  }{\sum\limits_{x_j\in\xone}  \exp\left( \frac{ -(\alpha y-x_j)^2 }{2 \alpha  \sigma^2} \right)  },\quad \alpha = \frac{1}{1+2\sigma^2\nu}.
\end{align}
Observe that (\ref{eq:BitLLR_MB}) differs from (\ref{eq:BitLLR_uniform}) only in the scaling factor $\alpha$. This means that to include the effect of $\Px$ during demapping, one should basically take a conventional demapper designed for uniform distribution, and scale its inputs by a constant, i.e. one does not need to implement a new demapper.

\subsubsection*{Decoding} Recall that the shaping bits do not carry  additional information, and their values are unknown to the receiver. During decoding of the last bit-level, the decoder should therefore treat the shaping bits in $\mathcal{S}$ as unknown bits, which may be discarded after the decoding is completed. Since the shaping bits are transmitted in the most reliable polar sub-channels (with relatively large indices), the decoder can also perform an early termination, as soon as all message bits are decoded. This can reduce the decoding latency. Alternatively the receiver can also use the shaping bits as an additional error detection mechanism similar to \cite{shaped_polar}.

\begin{remark}
We highlight that both modifications at the receiver are simple, and do not cause a significant increase in the receiver complexity. Recall that the transmitter has also the same order of complexity as a conventional MLC transmitter. As a result, the proposed scheme has an overall complexity comparable to a conventional MLC scheme. 
\end{remark}

\section{Performance Evaluation}
\label{sec:PerfEval}

In this section, we evaluate the block error rate (BLER) performance of the proposed scheme on AWGN channels using Monte-Carlo simulations. We perform simulations with $n_c = 256$ and $s=84$ shaping bits ($R_s\approx 1/3$) for $m\in\{2,3,4\}$, resulting in MB distributions with $\nu_2=0.171$, $\nu_3=0.041$ and $\nu_4=0.010$, respectively. At the transmitter, we use an SC decoder (which is equivalent to an SCL decoder with list size $\LE=1$) with min-sum approximation for sign-bit shaping. As an outer code, we use a 4-bit CRC with polynomial 0x3, which is attached at the end of the message sequence and transmitted in the sign bit-level. We use the polar sequence $\mathbf{Q}$ from \cite{chan_code5G} to construct the sets $\mathcal{S}$ (for the sign bit-level only), $\mathcal{I}$ and $\mathcal{F}$ for each bit-level. At the receiver, we use a list multi-stage decoder with $\LD=8$ (as described in  Sec. \ref{sec:MLC}), and pick the codeword with the best PM from the list that satisfies the cyclic redundancy check.
Table \ref{tab:params} contains the number of message bits at each bit-level ($k_i$, $ i\in\{1,\cdots,m \}$), which is obtained by numerical search to obtain good BLER results. We use the following three references to compare with our results:
\begin{itemize}
	\item As first reference, we use conventional uniform MLC with successive demapping, where we use the same setup as above except for $s=0$ (no shaping). The choice of $k_i$ is also given in Table \ref{tab:params}.
	\item As second reference, we use BICM transmission with Gray labeling. We design codes of length $m n_c$ with the same $\mathbf{Q}$, apply rate matching (when necessary, e.g. for $8$-ASK) as it is done in 5G NR and use the triangular channel interleaver according to \cite{chan_code5G}. We use the same 4-bit CRC as an outer code as above. At the receiver, we use parallel demapping and SCL decoding with $\LD=8$.
	\item As last reference, we evaluate Gallager's random coding (achievability) bound (RCB) \cite{gallager1968information} for Gaussian inputs with BLER given as $2^{-n_c E_\text{r}}$, where
	\begin{align}
	E_\text{r} = \max_{0\leq\rho\leq 1}\left\lbrace  \frac{\rho}{2} \cdot \log_2 \left( 1+\frac{\text{SNR}}{1+\rho}\right) - \frac{\rho\cdot k}{n_c} \right\rbrace.  
	\end{align}
\end{itemize}

\begin{figure}
	\centering
\includegraphics{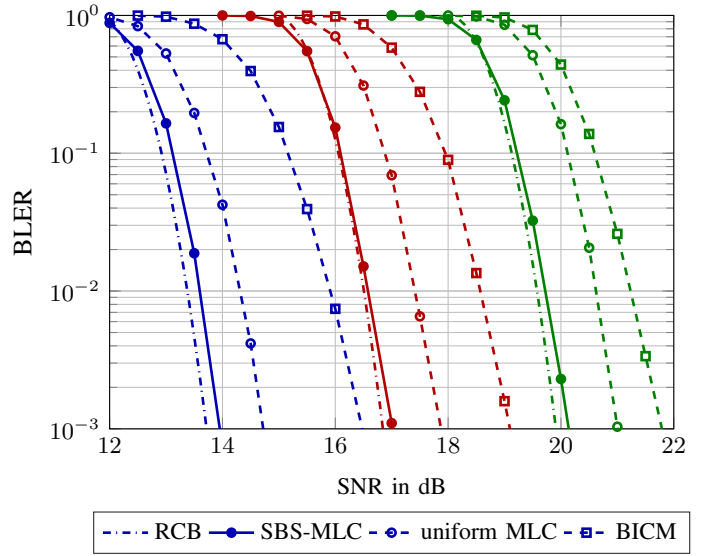}	
	\caption{BLER performance for 16-ASK ($m=4$) with uniform MLC, SBS-MLC, and BICM for rates $R = \{2, 2.5, 3\}$ bit/use.}
	\label{fig:BLERs4}
\end{figure}

Fig. \ref{fig:BLERs4} plots the BLER results for $16$-ASK ($m=4$). We observe that MLC shows improvements compared to BICM, as no demapping loss occurs. Moreover, using the proposed shaping approach, the performance can be further improved and the BLER curves approach the RCB. Recall that we use an SC decoder for sign-bit shaping, i.e. the complexity of the MLC approach and the SBS-MLC approach are similar. 

\begin{figure*}
	\centering
\includegraphics{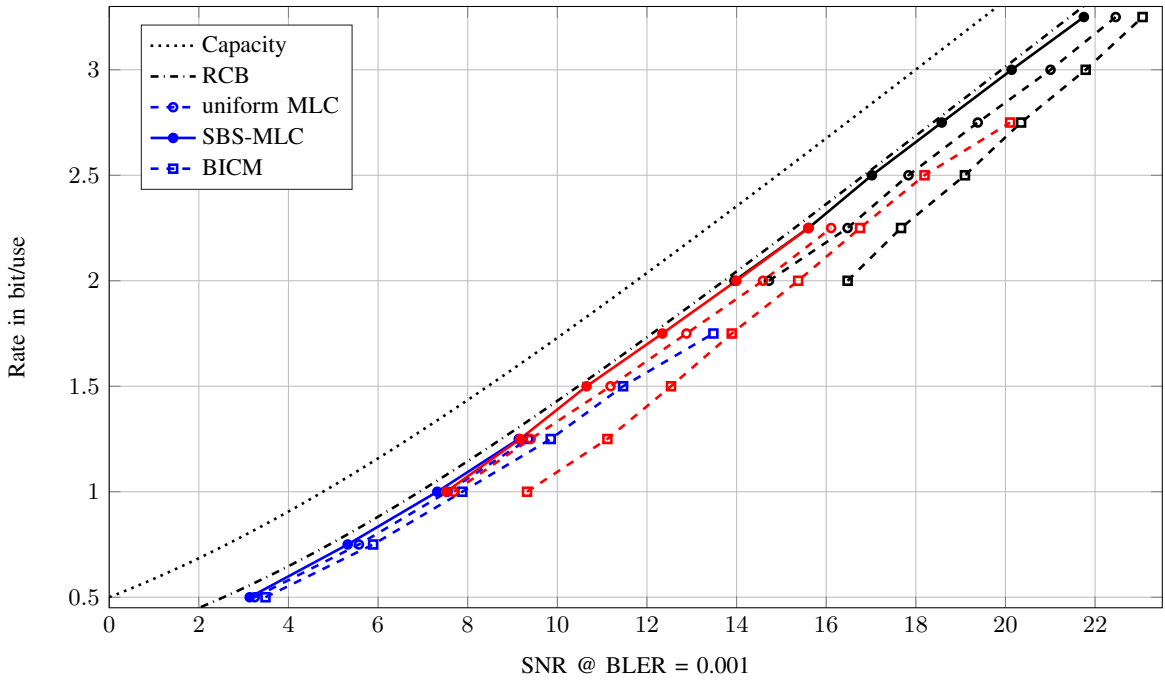}	
	\caption{Required SNR to achieve a target BLER 0.001 of the presented codes with 4-ASK (blue), 8-ASK (red) and 16-ASK (black).   }
	\label{fig:AchRateBLERS}
\end{figure*}

\begin{figure}
	\centering
	\includegraphics{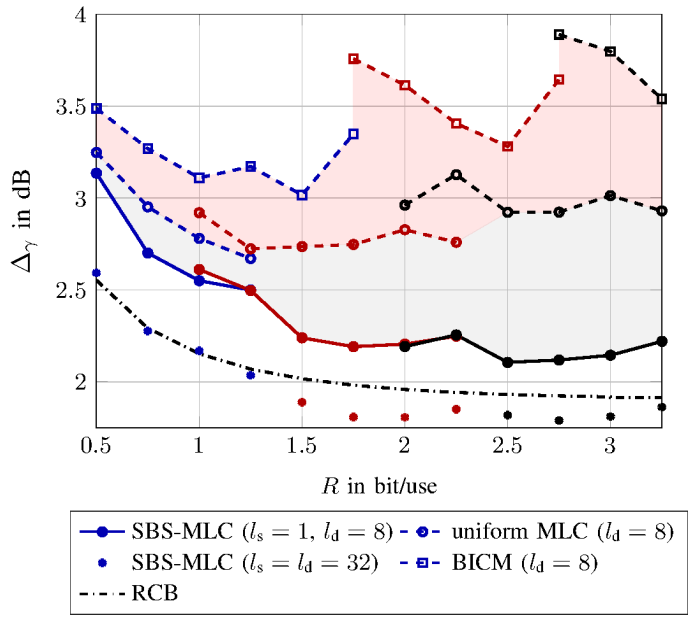}	
	\caption{SNR gap to capacity at a target BLER  0.001 with 4-ASK (blue), 8-ASK (red) and 16-ASK (black).}
	\label{fig:GaptoShannonL8}
\end{figure}

For a fair comparison between different transmission schemes, we also show results with different modulation orders in Fig. \ref{fig:AchRateBLERS}. We observe that the proposed scheme outperforms BICM and uniform MLC also for other modulation orders, and closely approaches RCB for all SNRs. To better visualize the gains in SNR, we plot in Fig. \ref{fig:GaptoShannonL8} the gap to channel capacity $\Delta_{\gamma}$ at each rate, corresponding to the horizontal distance of the points in Fig. \ref{fig:AchRateBLERS} to the Shannon limit. In this figure, the pink area can be seen as the demapping loss of BICM, and the green area can be interpreted as the shaping loss due to uniform signaling. We observe that especially at high rates, both demapping and shaping losses are significant. For example at $R=3$ bit/use, SBS-MLC performs 1.65dB better than BICM, and 0.88dB better than uniform MLC, i.e. for this example approximately $42\%$ of $\Delta_{\gamma}$ for BICM can be compensated by SBS-MLC, where $20\%$ of the gain comes from successive demapping, and $22\%$ from signal shaping.

\begin{figure}
	\centering
	\includegraphics{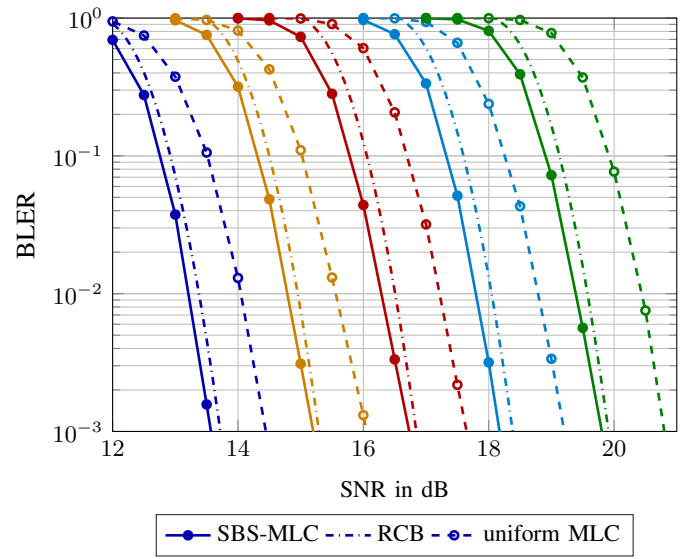}	
	\caption{BLER performance for 16-ASK ($m=4$) with SBS-MLC for rates $R = \{2, 2.25, 2.5, 2.75, 3\}$ bit/use with $\LE=\LD=32$, and for uniform MLC with $\LD=32$.}
	\label{fig:BLERm4_L32}
\end{figure}

Finally, we plot the BLER performance of the same codes for list sizes $\LD=32$ and $\LE=32$ in Fig. \ref{fig:BLERm4_L32}, and their gaps to capacity in Fig. \ref{fig:GaptoShannonL8}. As expected, there is an overall improvement due to the larger list size. Moreover, we observe that in this case the proposed SBS-MLC shows BLER performance even better than the RCB. 

\begin{table}
\centering
\caption{Simulation Parameters}
\label{tab:params}

	\begin{tabular}{cp{0.25cm}ccc|cccc|cc}
		&$R$ & $m$ &$n_c$& $k$ & $k_1$ & $k_2$ & $k_3$ & $k_4$ &CRC &$s$ \\   
		\hline
		\hline
		\multirow{14}{*}{\rot{SBS-MLC}}
		&0.5 	&2 & 256 & 128 & 21 	& 107 & - & - &4 &84 \\   
		&0.75 	&2 & 256 & 192 & 54 	& 138 & - & - &4 &84 \\   
		&1 		&2 & 256 & 256 & 100 	& 156 & - & - &4 &84 \\
		&1.25	&2 & 256 & 320 & 156 	& 164 & - & - &4 &84 \\
		\cline{2-11}
		&1 		&3 & 256 & 256 & 2 		& 100 & 154 & - &4 &84 \\   
		&1.25 	&3 & 256 & 320 & 10 	& 148 & 162 & - &4 &84 \\   
		&1.5 	&3 & 256 & 384 & 21 	& 195 & 168 & - &4 &84 \\   
		&1.75 	&3 & 256 & 448 & 54 	& 226 & 168 & - &4 &84 \\   
		&2 		&3 & 256 & 512 & 100 	& 244 & 168 & - &4 &84 \\   
		&2.25 	&3 & 256 & 576 & 156 	& 252 & 168 & - &4 &84 \\   
		\cline{2-11}
		&2 		&4 & 256 & 512 & 2 		& 100 & 246 & 164 &4 &84 \\   
		&2.25 	&4 & 256 & 576 & 10 	& 148 & 250 & 168 &4 &84 \\   
		&2.5 	&4 & 256 & 640 & 21 	& 195 & 256 & 168 &4 &84 \\   
		&2.75 	&4 & 256 & 704 & 54 	& 226 & 256 & 168 &4 &84 \\   
		&3 		&4 & 256 & 768 & 100 	& 244 & 256 & 168 &4 &84 \\   
		&3.25	&4 & 256 & 832 & 156 	& 252 & 256 & 168 &4 &84 \\   		
		\hline
		\hline
		\multirow{14}{*}{\rot{uniform MLC}}
&0.5 	&2 & 256 & 128 & 14 & 114 & - & - &4 &- \\   
&0.75 	&2 & 256 & 192 & 22 & 170 & - & - &4 &- \\   
&1 		&2 & 256 & 256 & 50 & 206 & - & - &4 &- \\
&1.25	&2 & 256 & 320 & 84 & 236 & - & - &4 &- \\   
\cline{2-11}
&1 		&3 & 256 & 256 & 2 	& 52  & 202 & - &4 &- \\   
&1.25 	&3 & 256 & 320 & 6 	& 80  & 234 & - &4 &- \\   
&1.5	&3 & 256 & 384 & 12 & 126 & 246 & - &4 &- \\   
&1.75 	&3 & 256 & 448 & 22 & 176 & 250 & - &4 &- \\   
&2		&3 & 256 & 512 & 50 & 212 & 250 & - &4 &- \\
&2.25 	&3 & 256 & 576 & 84 & 240 & 252 & - &4 &- \\     
\cline{2-11}
&2 		&4 & 256 & 512 & 2  & 52  & 208 & 250 &4 &- \\ 
&2.25 	&4 & 256 & 576 & 6  & 80  & 238 & 252 &4 &- \\   
&2.5 	&4 & 256 & 640 & 10 & 130 & 248 & 252 &4 &- \\   
&2.75	&4 & 256 & 704 & 22 & 176 & 254 & 252 &4 &- \\   
&3 		&4 & 256 & 768 & 50 & 212 & 254 & 252 &4 &- \\
&3.25	&4 & 256 & 832 & 84& 240 & 256 & 252 &4 &- \\   		   
	\end{tabular}

\end{table}

\section{Conclusion}
\label{sec:Conc}

In this work, we studied higher order modulation with signal shaping for AWGN channels. 
Through theoretical evaluations, we first demonstrated that capacity of the AWGN channels can be approached with successive demapping, even if the symbol distribution is not optimized for the operating SNR. This is in general not the case for approaches based on parallel demapping such as BICM that require the symbol distribution to be optimized for the operating SNR. We further showed that shaping only the sign bit of ASK symbols with natural labeling using a fixed shaping rate $R_s$ is sufficient to compensate for the shaping loss over a wide SNR range. Based on these facts, we elaborated on the choice of the parameters, and designed a sign-bit shaped multi-level polar coding scheme, which roughly has the same complexity as conventional uniform MLC. The proposed scheme does not require any additional shaping decoders at the receiving side, and hence the receiver is similar to a conventional MLC receiver. By numerical simulations we have demonstrated that the proposed scheme shows superior BLER performance compared to BICM and MLC approaches with uniform symbol distributions, and approach RCB for $n_c=256$ channel uses. We also show that by allowing a higher complexity, the RCB bound can be outperformed.

\bibliographystyle{IEEEtran}
\bibliography{IEEEabrv,mybibfile}

\begin{thebibliography}{10}
\providecommand{\url}[1]{#1}
\csname url@samestyle\endcsname
\providecommand{\newblock}{\relax}
\providecommand{\bibinfo}[2]{#2}
\providecommand{\BIBentrySTDinterwordspacing}{\spaceskip=0pt\relax}
\providecommand{\BIBentryALTinterwordstretchfactor}{4}
\providecommand{\BIBentryALTinterwordspacing}{\spaceskip=\fontdimen2\font plus
\BIBentryALTinterwordstretchfactor\fontdimen3\font minus
  \fontdimen4\font\relax}
\providecommand{\BIBforeignlanguage}[2]{{%
\expandafter\ifx\csname l@#1\endcsname\relax
\typeout{** WARNING: IEEEtran.bst: No hyphenation pattern has been}%
\typeout{** loaded for the language `#1'. Using the pattern for}%
\typeout{** the default language instead.}%
\else
\language=\csname l@#1\endcsname
\fi
#2}}
\providecommand{\BIBdecl}{\relax}
\BIBdecl

\bibitem{Arikan09}
E.~Ar{\i}kan, ``Channel polarization: A method for constructing
  capacity-achieving codes for symmetric binary-input memoryless channels,''
  \emph{{IEEE} Trans. Inf. Theory}, vol.~55, no.~7, pp. 3051--3073, Jul. 2009.

\bibitem{Tal15}
I.~Tal and A.~Vardy, ``List decoding of polar codes,'' \emph{{IEEE} Trans. Inf.
  Theory}, vol.~61, no.~5, pp. 2213--2226, May 2015.

\bibitem{chan_code5G}
\emph{3GPP TS 38.212 Technical Specification Group Radio Access Network, NR,
  Multiplexing and Channel Coding}, 2017.

\bibitem{hussami2009performance}
N.~Hussami, S.~B. Korada, and R.~Urbanke, ``Performance of polar codes for
  channel and source coding,'' in \emph{Proc. {IEEE} Int. Symp. Inf. Theory
  (ISIT)}, 2009, pp. 1488--1492.

\bibitem{korada2010polar}
S.~B. Korada and R.~L. Urbanke, ``Polar codes are optimal for lossy source
  coding,'' \emph{{IEEE} Trans. Inf. Theory}, vol.~56, no.~4, pp. 1751--1768,
  2010.

\bibitem{goela2014polar}
N.~Goela, E.~Abbe, and M.~Gastpar, ``Polar codes for broadcast channels,''
  \emph{{IEEE} Trans. Inf. Theory}, vol.~61, no.~2, pp. 758--782, 2014.

\bibitem{honda2013polar}
J.~Honda and H.~Yamamoto, ``Polar coding without alphabet extension for
  asymmetric models,'' \emph{{IEEE} Trans. Inf. Theory}, vol.~59, no.~12, pp.
  7829--7838, 2013.

\bibitem{andersson2010nested}
M.~Andersson, V.~Rathi, R.~Thobaben, J.~Kliewer, and M.~Skoglund, ``Nested
  polar codes for wiretap and relay channels,'' \emph{{IEEE} Commun. Lett.},
  vol.~14, no.~8, pp. 752--754, 2010.

\bibitem{6034749}
H.~{Mahdavifar} and A.~{Vardy}, ``Achieving the secrecy capacity of wiretap
  channels using polar codes,'' \emph{{IEEE} Trans. Inf. Theory}, vol.~57,
  no.~10, pp. 6428--6443, Oct. 2011.

\bibitem{shaped_polar}
O.~\.{I}\c{s}can, R.~B{\"o}hnke, and W.~Xu, ``Shaped polar codes for higher
  order modulation,'' \emph{{IEEE} Commun. Lett.}, vol.~22, no.~2, pp.
  252--255, 2018.

\bibitem{8691487}
O.~{G\"unl\"u}, O.~{\.I}{\c{s}}can, V.~{Sidorenko}, and G.~{Kramer}, ``Code
  constructions for physical unclonable functions and biometric secrecy
  systems,'' \emph{{IEEE} Trans. Inf. Forensics Security}, vol.~14, no.~11, pp.
  2848--2858, Nov. 2019.

\bibitem{seidl2013polar}
M.~Seidl, A.~Schenk, C.~Stierstorfer, and J.~B. Huber, ``Polar-coded
  modulation,'' \emph{{IEEE} Trans. Commun.}, vol.~61, no.~10, pp. 4108--4119,
  Oct. 2013.

\bibitem{imai1977new}
H.~Imai and S.~Hirakawa, ``A new multilevel coding method using
  error-correcting codes,'' \emph{{IEEE} Trans. Inf. Theory}, vol.~23, no.~3,
  pp. 371--377, 1977.

\bibitem{wachsmann1999multilevel}
U.~Wachsmann, R.~F. Fischer, and J.~B. Huber, ``Multilevel codes: Theoretical
  concepts and practical design rules,'' \emph{{IEEE} Trans. Inf. Theory},
  vol.~45, no.~5, pp. 1361--1391, 1999.

\bibitem{caire1998bit}
G.~Caire, G.~Taricco, and E.~Biglieri, ``Bit-interleaved coded modulation,''
  \emph{{IEEE} Trans. Inf. Theory}, vol.~44, no.~3, pp. 927--946, 1998.

\bibitem{bocherer2017efficient}
G.~B\"ocherer, T.~Prinz, P.~Yuan, and F.~Steiner, ``Efficient polar code
  construction for higher-order modulation,'' in \emph{Proc. {IEEE} Wireless
  Comm. and Netw. Conf. (WCNC)}, Mar. 2017, pp. 1--6.

\bibitem{mahdavifar2015polar}
H.~Mahdavifar, M.~El-Khamy, J.~Lee, and I.~Kang, ``Polar coding for
  bit-interleaved coded modulation,'' \emph{{IEEE} Trans. Veh. Technol.},
  vol.~65, no.~5, pp. 3115--3127, 2015.

\bibitem{shin2012mapping}
D.-M. Shin, S.-C. Lim, and K.~Yang, ``Mapping selection and code construction
  for $2^m$-ary polar-coded modulation,'' \emph{{IEEE} Commun. Lett.}, vol.~16,
  no.~6, pp. 905--908, 2012.

\bibitem{chen2013efficient}
K.~Chen, K.~Niu, and J.-R. Lin, ``An efficient design of bit-interleaved polar
  coded modulation,'' in \emph{Proc. {IEEE} Int. Symp. Pers., Indoor and Mob.
  Rad. Comm. (PIMRC)}, Sep. 2013, pp. 693--697.

\bibitem{kschischang1993optimal}
F.~R. Kschischang and S.~Pasupathy, ``Optimal nonuniform signaling for
  {Gaussian} channels,'' \emph{{IEEE} Trans. Inf. Theory}, vol.~39, no.~3, pp.
  913--929, 1993.

\bibitem{mondelli2018achieve}
M.~Mondelli, S.~H. Hassani, and R.~L. Urbanke, ``How to achieve the capacity of
  asymmetric channels,'' \emph{{IEEE} Trans. Inf. Theory}, vol.~64, no.~5, pp.
  3371--3393, 2018.

\bibitem{icscan2019probabilistic}
O.~{\.I}{\c{s}}can, R.~B{\"o}hnke, and W.~Xu, ``Probabilistic shaping using
  {5G} new radio polar codes,'' \emph{{IEEE} Access}, vol.~7, pp.
  22\,579--22\,587, 2019.

\bibitem{lnt_hwdu}
T.~Prinz, P.~Yuan, G.~B\"ocherer, F.~Steiner, O.~\.{I}\c{s}can, R.~B\"ohnke,
  and W.~Xu, ``Polar coded probabilistic amplitude shaping for short packets,''
  in \emph{Proc. {IEEE} Int. Workshop Signal Process. Advances Wireless Commun.
  (SPAWC)}, Jul. 2017, pp. 83--87.

\bibitem{bocherer2015bandwidth}
G.~B{\"o}cherer, F.~Steiner, and P.~Schulte, ``Bandwidth efficient and
  rate-matched low-density parity-check coded modulation,'' \emph{{IEEE} Trans.
  Commun.}, vol.~63, no.~12, pp. 4651--4665, 2015.

\bibitem{icscan2018probabilistically}
O.~{\.I}{\c{s}}can, R.~B{\"o}hnke, and W.~Xu, ``Probabilistically shaped
  multi-level coding with polar codes for fading channels,'' in \emph{Proc.
  {IEEE} Global Commun. Conf. Wkshps (GC Wkshps)}, Dec. 2018, pp. 1--5.

\bibitem{8492454}
L.~{Liu}, Y.~{Yan}, C.~{Ling}, and X.~{Wu}, ``Construction of
  capacity-achieving lattice codes: Polar lattices,'' \emph{{IEEE} Trans.
  Commun.}, vol.~67, no.~2, pp. 915--928, 2019.

\bibitem{forney1992trellis}
G.~Forney, ``Trellis shaping,'' \emph{{IEEE} Trans. Inf. Theory}, vol.~38,
  no.~2, pp. 281--300, 1992.

\bibitem{smith2012pragmatic}
B.~P. Smith and F.~R. Kschischang, ``A pragmatic coded modulation scheme for
  high-spectral-efficiency fiber-optic communications,'' \emph{J. Lightw.
  Technol.}, vol.~30, no.~13, pp. 2047--2053, 2012.

\bibitem{gallager1968information}
R.~G. Gallager, \emph{Information Theory and Reliable Communication}.\hskip 1em
  plus 0.5em minus 0.4em\relax New York, NY, USA: Wiley, 1968.

\bibitem{bocherer2014achievable}
G.~B{\"o}cherer, ``Achievable rates for shaped bit-metric decoding,''
  \emph{arXiv preprint arXiv:1410.8075}, 2014.

\bibitem{mori2009performance}
R.~Mori and T.~Tanaka, ``Performance of polar codes with the construction using
  density evolution,'' \emph{{IEEE} Commun. Lett.}, vol.~13, no.~7, 2009.

\bibitem{Tal13}
I.~Tal and A.~Vardy, ``How to construct polar codes,'' \emph{{IEEE} Trans. Inf.
  Theory}, vol.~59, no.~10, pp. 6562--6582, Oct. 2013.

\bibitem{trifonov2012efficient}
P.~Trifonov, ``Efficient design and decoding of polar codes,'' \emph{{IEEE}
  Trans. Commun.}, vol.~60, no.~11, pp. 3221--3227, 2012.

\bibitem{dettmar1992new}
U.~Dettmar, J.~Portugheis, and H.~Hentsch, ``New multistage decoding
  algorithm,'' \emph{Electron. Lett.}, vol.~28, no.~7, pp. 635--636, 1992.

\bibitem{B-Stimming15}
A.~Balatsoukas-Stimming, M.~B. Parizi, and A.~Burg, ``{LLR}-based successive
  cancellation list decoding of polar codes,'' \emph{IEEE Trans. on Sig.
  Proc.}, vol.~63, no.~19, pp. 5165--5179, 2015.

\bibitem{tse2005fundamentals}
D.~Tse and P.~Viswanath, \emph{Fundamentals of Wireless Communication}.\hskip
  1em plus 0.5em minus 0.4em\relax New York, NY, USA: Cambridge University
  Press, 2005.

\end{thebibliography}

\end{document}